# Instabilities of Advection-Dominated Accretion Flows

Xingming CHEN

*Department of Astronomy and Astrophysics, Göteborg University
and Chalmers University of Technology, 412 96 Göteborg, Sweden*

## Abstract

Accretion disk instabilities are briefly reviewed. Some details are given to the short-wavelength thermal instabilities and the convective instabilities. Time-dependent calculations of two-dimensional advection-dominated accretion flows are presented.

## 1. Introduction

Accretion disk is central to our understanding of many diverse astronomical objects and phenomena. In particular, it is believed that disk instabilities are fundamentally responsible for some of the luminosity variations observed in these systems.

Various kinds of instabilities may arise in a rotating, shearing gas flow. For example, the MHD instability in a shearing flow (Balbus & Hawley 1991) is found to be able to survive and develop into nonlinear turbulence. In fact, an external magnetic field is not required to sustain the instability since magnetic field can be generated and maintained self-consistently through dynamo-effect within the flow (Brandenburg et al. 1995). The Balbus-Hawley instability is not considered here, instead, I model the small-scale turbulence by assuming that it introduces viscosity, which can be described by the $\alpha$ prescription (Shakura & Sunyaev 1973),

$$\nu = \frac{2}{3}\alpha c_s H, \tag{1}$$

where $c_s$ is the local sound speed and $H$ is the scale-height of the disk. The disk is also assumed to be axisymmetric and non-gravitating, therefore the global non-axisymmetric Papaloizou-Pringle instability (Papaloizou & Pringle 1984) and all the self-gravitating-related instabilities are not present here.

An $\alpha$-model accretion disk may experience the thermal and viscous instabilities (Piran 1978). They could be coherent global instabilities and are often applied to explain the dwarf nova outbursts and the X-ray transients (see related discussion in this volume and references within). This type of instability can be analysed conveniently in terms of the $\dot{M}(\Sigma)$ relation (where $\dot{M}$ is the mass accretion rate and $\Sigma$ is the surface density of the disk). Here I consider the case of accretion disks around black holes. The steady state solutions or the thermal equilibria of accretion disks can be described in a unified scheme (see Chen et al. 1995). Depending on the magnitude of the viscosity parameter $\alpha$, the cooling mechanisms, and the equation of states, there exist several different branches (see Fig. 1). On the $\dot{M}(\Sigma)$ plane, it can be shown that, if the slope is negative, then the disk is viscously unstable (Lynden-Bell & Pringle 1974). The thermal instability can be examined by comparing the cooling and heating rates near each equilibrium curve. One may consider a perturbation of $\dot{M}$ (increase $\dot{M}$ but keep $\Sigma$ unchanged), or effectively an increase of temperature, if the cooling rate exceeds the heating rate, then the gas cools down and the original thermal equilibrium is restored. The disk is therefore thermally stable. On the other hand, if the heating rate exceeds the cooling rate, then the gas heats up further and an instability occurs.

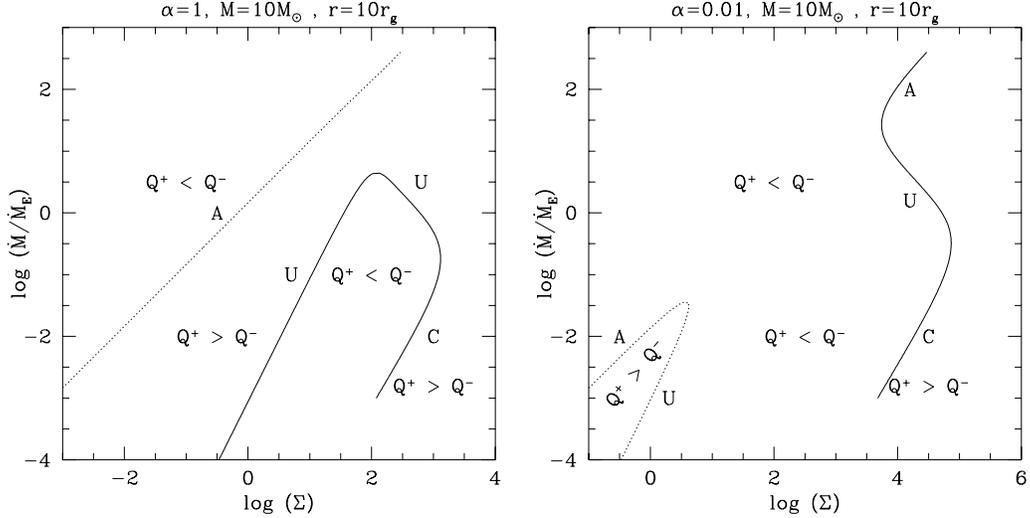

Fig. 1. The thermal equilibria of accretion disks. Here $\dot{M}_{\rm E} = 4\pi GM/\kappa c$ is the Eddington limit ($\kappa = 0.34$), $Q^+$ is the heating rate, and $Q^-$ is the cooling rate including both local radiative cooling and radial advective cooling. A: 'advection' and stable, U: 'unstable', and C: 'local cooling' and stable.

Thus, it is easily understood that for optically thick disks (large surface density), the thermal and viscous instabilities occur at the same time on the middle-branch of the **S**-shaped curve (or on the upper-branch of the upside down **U**-curve) when the disk is radiation pressure dominated and the radial advection is not important. In this case a limit-cycle type disk evolution may be resulted (Taam & Lin 1984) or a transition to an optically thin disk state may occur (Lightman & Eardley 1974). For optically thin local cooling dominated disks, on the other hand, the viscous mode is stable but the thermal mode is always unstable. Therefore, a stable optically thin local cooling dominated accretion disk may not exist, it may either collapse to an optically thick disk or heat up catastrophically without bound. Finally but most importantly, the advection dominated disks are both thermally and viscously stable independent on the optically depth.

The thermal and viscous instabilities are probably the most violent global instabilities an accretion disk around black hole may suffer. Under such instabilities, a disk will most probably be destroyed entirely, especially in the case of optically thin disks. Accretion disk models which do not have such type of instabilities include advection-dominated disks and gas pressure dominated optically thick disks. In the following, I discuss the less severe instabilities which may exist in advection-dominated accretion disks.

## 2. Short-Wavelength Thermal Instability

The above analysis for thermal instability, however, can only be applied to the long-wavelength perturbations. To examine the short-wavelength instability, a dispersion relation equation needs to be solved. The dispersion relation can be derived from the time-dependent disk equations by assuming the perturbation (of the steady state solution), $\delta X$, of each physical quantity, $X$, has the following functional form with respect to time ($t$) and space (r):

$$\delta X / X \propto e^{\xi t - ikr}, \qquad (2)$$

where $k$ is the wavenumber of the modes, $k = 2\pi/\lambda$, and $\xi$ is, in general, complex. Specifically, $\tau = \text{Re}(\xi)$ represents the growth or damping rate depending upon whether $\tau$ is positive or negative, and $\omega = \text{Im}(\xi)$ represents the frequency of the mode.

I here consider only the thermal mode instability. In a recent work, Kato, Abramowicz, & Chen (1996) demonstrated, analytically, that short-wavelength thermal mode instability is still present in advection-dominated accretion disks. The unstable short-wavelength thermal mode has been obtained numerically in Chen & Taam (1993). However, detailed discussion has not been made there.

By solving the dispersion relation of a transonic black hole accretion disk model (see Chen & Taam 1993), it is seen that the short-wavelength ($\lambda \lesssim 6H$) thermal instability is a traveling mode with frequencies in the order of the local dynamical time-scale. Figure 2 shows the growth rate and the oscillation frequency of the thermal mode instability for two cases with $\dot{M} = 16\dot{M}_\text{E}$ and $\dot{M} = 160\dot{M}_\text{E}$, where $\dot{M}_\text{E} = 4\pi GM/\kappa c$ is the Eddington limit and $\kappa = 0.34$. The disk is optically thick with the latter one fully advection dominated. Because of the nature of the propagation, the instability is expected to be weak globally and exist only in the inner regions near the black hole. Local properties of the short-wavelength thermal mode instability suggest that its non-linear behavior may be similar to that of the inertial-acoustic mode instability discussed in Chen & Taam (1995). Preliminary results of Manmoto et al. (1995, this volume) shows that a perturbation indeed grows due the short-wavelength thermal mode instability, but then it is advected into the black hole.

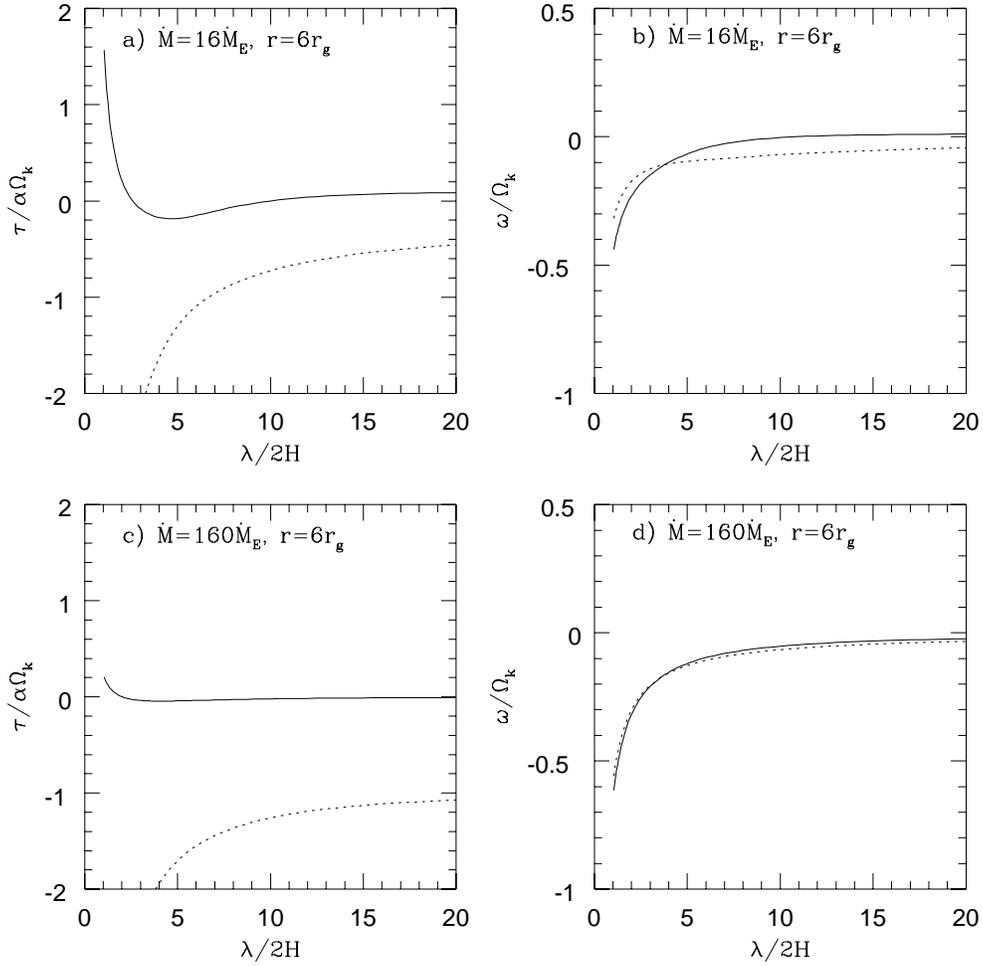

Fig. 2. The growth rate and the oscillation frequency of the thermal mode (solid line, the dotted line represents the viscous mode). Note the increase of the thermal mode growth rate for short-wavelength ($\lambda \lesssim 6H$), and also the corresponding high frequencies ($\omega \sim \Omega_K$).

## 3. Convective Instability: Time-Dependent Simulations

In an advection-dominated disk, the specific entropy of the flow increases inwards, therefore, it is likely that the convective instability is present. For a static rotating fluid, the convective stability condition is the Høiland criterion which states that the fluid is stable against local, axisymmetric, adiabatic perturbations if and only if the following two conditions are satisfied (Tassoul 1978):

$$\frac{1}{r^3}\frac{\partial \ell^2}{\partial r} - \left(\frac{\partial T}{\partial p}\right)_S \nabla p \cdot \nabla S > 0, \qquad (3a)$$

$$-\frac{1}{\rho}\frac{\partial p}{\partial z}\left(\frac{\partial \ell^2}{\partial r}\frac{\partial S}{\partial z} - \frac{\partial \ell^2}{\partial z}\frac{\partial S}{\partial r}\right) > 0. \qquad (3b)$$

Here $(r, z)$ is the radial and vertical coordinates in a cylindrical coordinates, $\ell$ is the specific angular momentum, $p$ is the pressure, and $S$ is the specific entropy. Note that the first terms in criterion (3a) is the epicyclic frequency $\kappa^2$ and this criterion reduces to that of Narayan & Yi (1994) for one-dimensional gas pressure dominated disks.

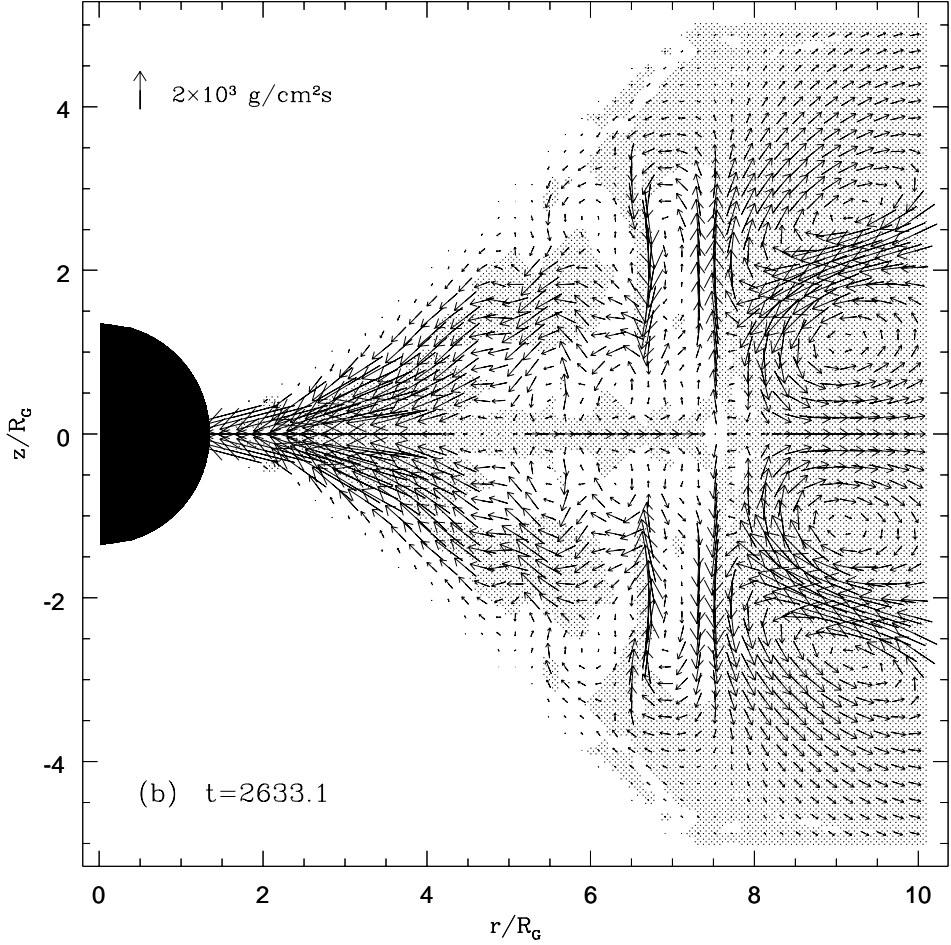

Fig. 3. The momentum vector field of advection-dominated flows. The regions which are convectively unstable according to Høiland criterion (3ab) are indicated as grey areas. The time is in unit of $r_g/c$. This model has $\alpha = 0.001$ and the initial condition is a static torus.

We have performed two-dimensional hydrodynamical simulations for axisymmetrical advection dominated accretion flows. 2D numerical models of accretion disks around black holes have been studied in different situations. For example, Hawley, Smarr, & Wilson (1984) simulated the formation of gas pressure supported tori. Eggum, Coroniti, & Katz (1987) calculated radiative hydrodynamical models of optically thick accretion disks. Our detailed results have been reported elsewhere (Igumenshchev, Chen, & Abramowicz 1996). A typical flow pattern (momentum vector field) is shown in Figure 3. Here, we have used a viscosity prescription $\nu = (2/3)\alpha c_s^2/\Omega_K$, and $\alpha = 0.001$. The grey area represents the convective unstable regions according to the Høiland criterion (3ab). We see a very efficient mixing of matter. Further analysis shows that this process of meridional circulations and vortices transports angular momentum outwards and thus the effective $\alpha$ becomes larger.

## 4. Summary

Advection dominated flows subject to convective instability. By time dependent simulations, we showed that this process transports angular momentum outwards. However, how important does it effect the dynamics of the flow is yet to be examined. There were some numerical simulations which showed that convection transports angular momentum inwards (e.g. Stones & Balbus 1995). The difference may be caused by the different physical conditions considered. For example, in our simulation, an $\alpha$-viscosity is assumed whereas in Stones & Balbus (1995) no viscosity is included.

In advection dominated flows, there is short-wavelength thermal instability. This type of instability is expected to be weak and localized and it may be termed as "flicker-type" instability (Kato et al. 1996). High resolution time-dependent calculation should be the future investigation.

I wish to thank Dr. Shoji Kato for support.